\documentstyle[12pt,epsfig]{article}

\def\Title#1{\begin{center} {\Large #1 } \end{center}}

\begin{document}


\pagestyle{empty}
\setcounter{page}{0}
\begin{flushright}
CDF/PUB/BOTTOM/PUBLIC/5200 \\
FERMILAB-Conf-00/007-E \\
Version 2.0 \\
\today 
\end{flushright}
\vskip 2.0cm

\begin{center}
{\Large\bf Brief Report from the Tevatron}
\end{center}
\vskip 1.0cm

\begin{center}
Manfred Paulini \\
{\em Lawrence Berkeley National Laboratory \\
Berkeley, California 94720, USA}
\vskip 0.5cm
(Representing the CDF and D\O\ Collaborations)
\end{center}

\vskip 1.5cm

\begin{abstract}
\noindent
We report on the $B$~physics prospects from the Fermilab Tevatron, 
summarizing the $B$~physics goals of the CDF and D\O\ experiments using
their upgraded detectors. We discuss the time schedule for completion of
the detector upgrades and summarize the current measurement of the
$CP$~violation parameter $\sin 2\beta$ at CDF.
\end{abstract}

\vskip 5.5cm

\begin{center}
Plenary talk presented at \\
XIX International
Symposium on Lepton and Photon Interactions at High Energies \\
Stanford University, August 9-14, 1999.
\end{center}

\newpage
\setcounter{page}{0}
\thispagestyle{empty}
\vbox{}
\newpage


\Title{Brief Report from the Tevatron}

\bigskip\bigskip


\begin{raggedright}  

{\it Manfred Paulini\index{author}{Paulini, Manfred}\\
Lawrence Berkeley National Laboratory \\
Berkeley, California 94720}
\bigskip\bigskip
\end{raggedright}

\section{Introduction}

It might appear surprising to include a report from the Fermilab Tevatron,
a proton-antiproton collider, in a session about ``Brief Reports from the
$B$~factories''. Does this mean the Tevatron would qualify as a $B$~factory? 
There are two advantages of studying $B$~physics at the Tevatron. First,
all $B$~hadrons are produced; not only charged and
neutral $B$~mesons as at the $B$~factories, but also 
$B_{\mbox{\scriptsize S}}^0$~mesons and $b$-baryons. The second advantage
is the $b$~quark production cross section, which is about 1~nb at the
$\Upsilon(4S)$ resonance while it is about 50~$\mu$b for $p\bar p$
collisions at $\sqrt{s}=1.8$~TeV. This
is an enormous cross section which is about 50,000 times
larger at the Tevatron than at the $B$~factories. It resulted in about
$5\times10^9$ $b\bar b$ quark pairs being produced during the 1992-1995
data taking period of the Tevatron, called Run\,I. 
To illustrate the enormous $b$~production rate at the Tevatron, we compare the
yield of fully reconstructed $B$~mesons between the CLEO experiment and
CDF. In a data sample of about 3000~pb$^{-1}$, CLEO reconstructs about 200
$B$~mesons decaying into $J/\psi K^+$~\cite{Jessop:1997} while CDF finds in a
sample of about 100~pb$^{-1}$ of data a signal of about 1000~$J/\psi K^+$
events with a good signal-to-background ratio~\cite{Abe:1999_1}. 

The goal of the $B$~factories is
to discover $CP$~violation in $B^0 \rightarrow J/\psi K^0_S$ decays.
CDF has already presented a measurement of $CP$~violation in
the $B$~meson system~\cite{Affolder:1999}, measuring the time-dependent
asymmetry in the yield of $J/\psi K^0_S$ events coming from a  
$B^0$ versus $\bar B^0$:
\begin{equation}
{\cal A}_{CP}(t) \equiv \frac{N(\bar{B}{^0}(t))-N(B^0(t)) }
                        {N(\bar{B}{^0}(t))+N(B^0(t)) } = 
                       \sin 2\beta \ \sin \Delta m_d t.
\end{equation}
This asymmetry is directly related to the $CP$~violation parameter 
$\sin 2\beta$. 

\subsection{CDF Measurement of \boldmath{$\sin 2\beta$}}

\begin{figure}[tb]
\begin{center}
\centerline{
\put(40,190){\large\bf (a)}
\put(250,190){\large\bf (b)}
\epsfysize=3.0in
\epsffile[5 5 530 530]{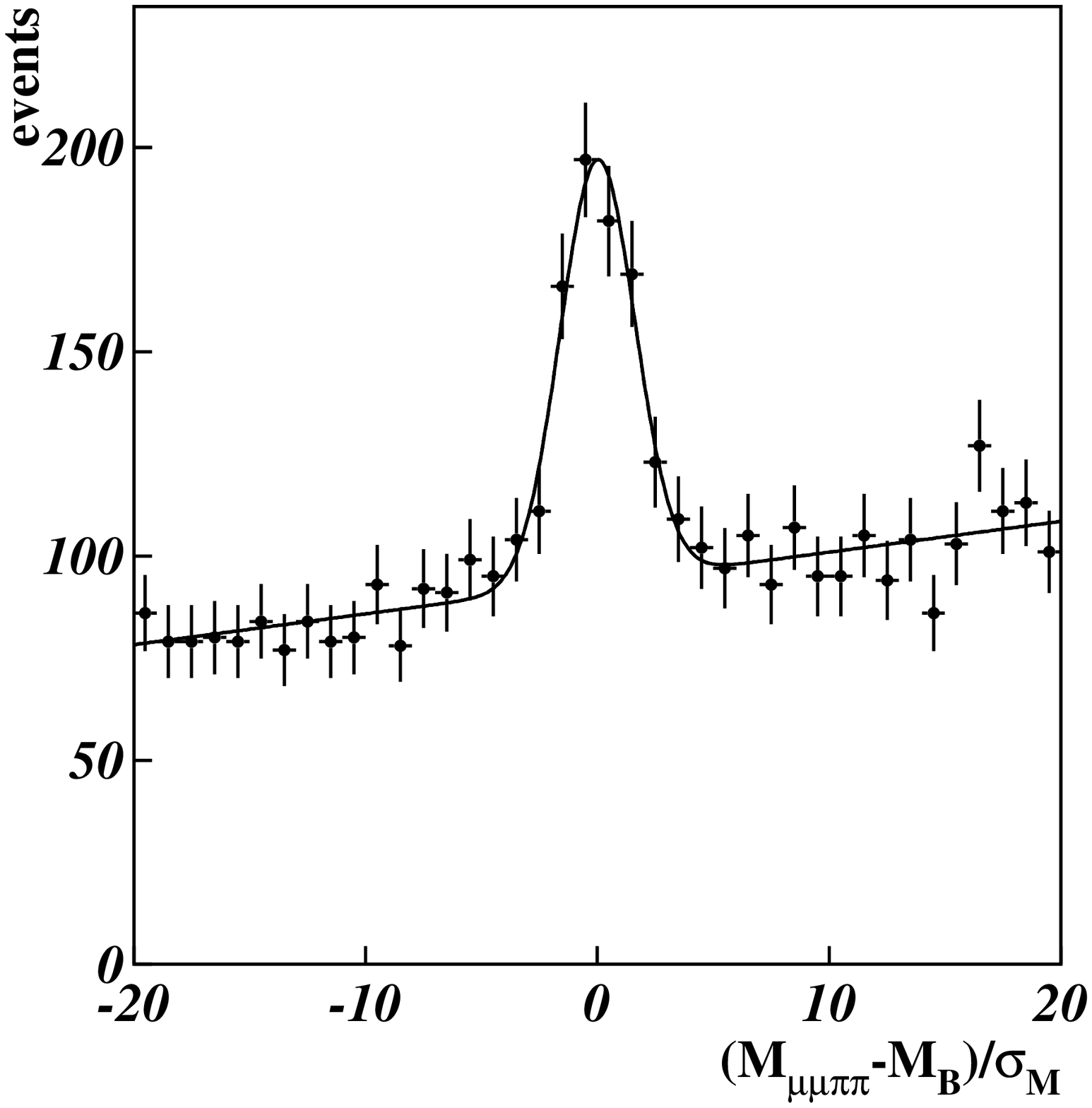}
\epsfysize=3.0in
\epsffile[5 10 545 530]{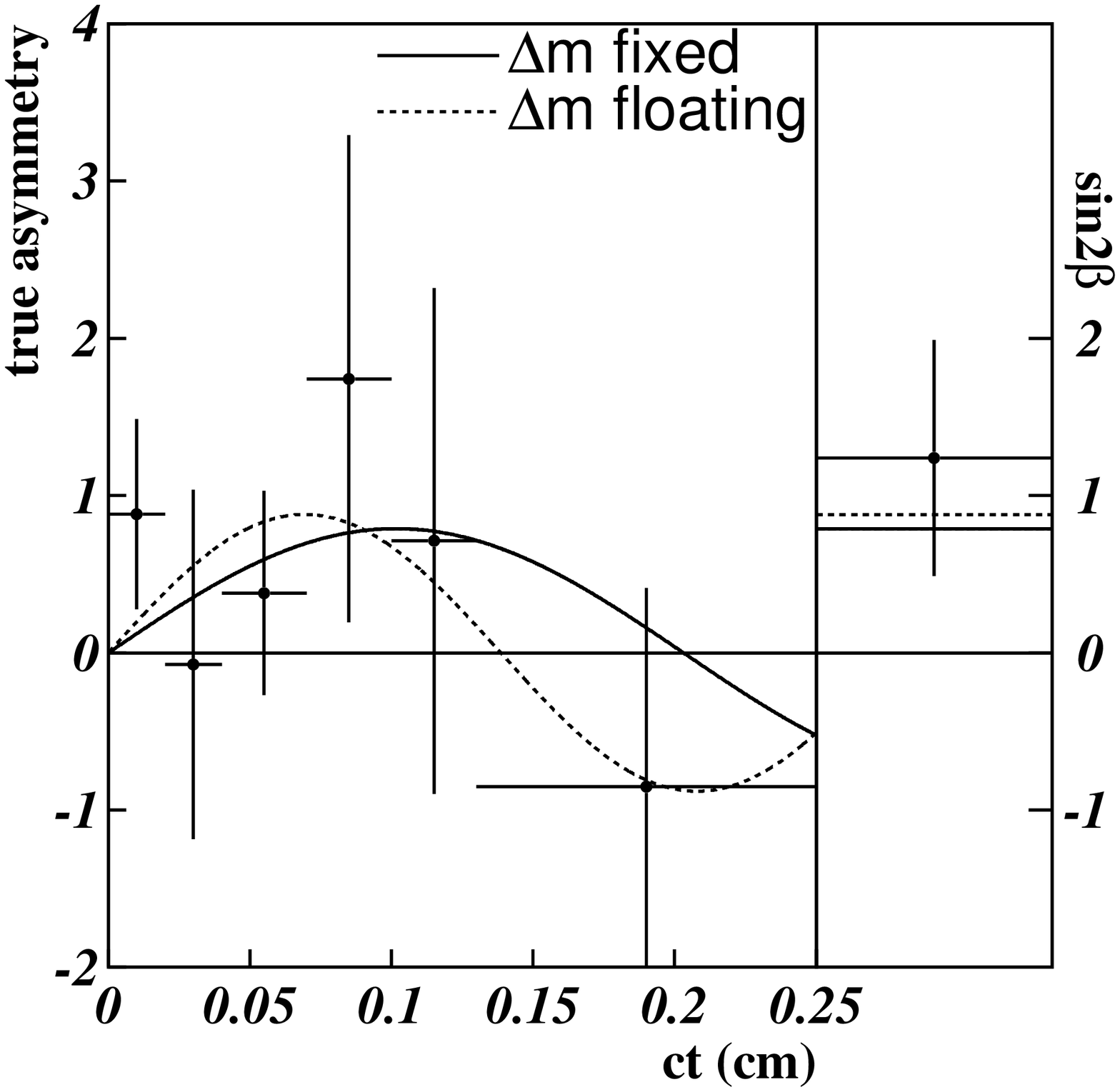}
}
\caption{
(a) Normalized mass distribution of $J/\psi K^0_S$ candidates. 
(b) True asymmetry $\sin 2\beta$ as a function 
of the reconstructed $J/\psi K^0_S$ proper decay length.
The events with low decay length resolution are shown separately on the right.}
\label{fig:cdfsin2b}
\end{center}
\end{figure}

Here, we briefly summarize CDF's initial measurement of $\sin 2\beta$.
Figure~\ref{fig:cdfsin2b}(a) shows the $J/\psi K^0_S$ yield at CDF, where
$395\pm31$ events have been identified.
This is currently the world's largest sample of fully reconstructed
$J/\psi K^0_S$ events.  
Measuring a $CP$~asymmetry requires knowing
whether the $J/\psi K^0_S$ originated from a $B^0$ or $\bar B^0$
meson. This is usually referred to as $B$~flavor
tagging. Several methods of $B$~flavor tagging exist. Some of them exploit
the other $B$~hadron in the event and search for a lepton from the
semileptonic decay of the other $B$~hadron or determine the net charge of
the jet produced by the other $b$~quark. These two methods are called lepton
tagging and jet charge tagging, respectively. The $B$~flavor can also be
determined by searching for pions which are produced through fragmentation
or $B^{**}$~mesons in correlation with the $B$~meson of interest. This
method is known as same side tagging. 

\begin{figure}[tb]
\vspace{.32in}
\begin{center}
\centerline{
\put(40,30){\large\bf (a)}
\put(400,40){\large\bf (b)}
\epsfysize=2.5in
\epsffile[5 5 530 490]{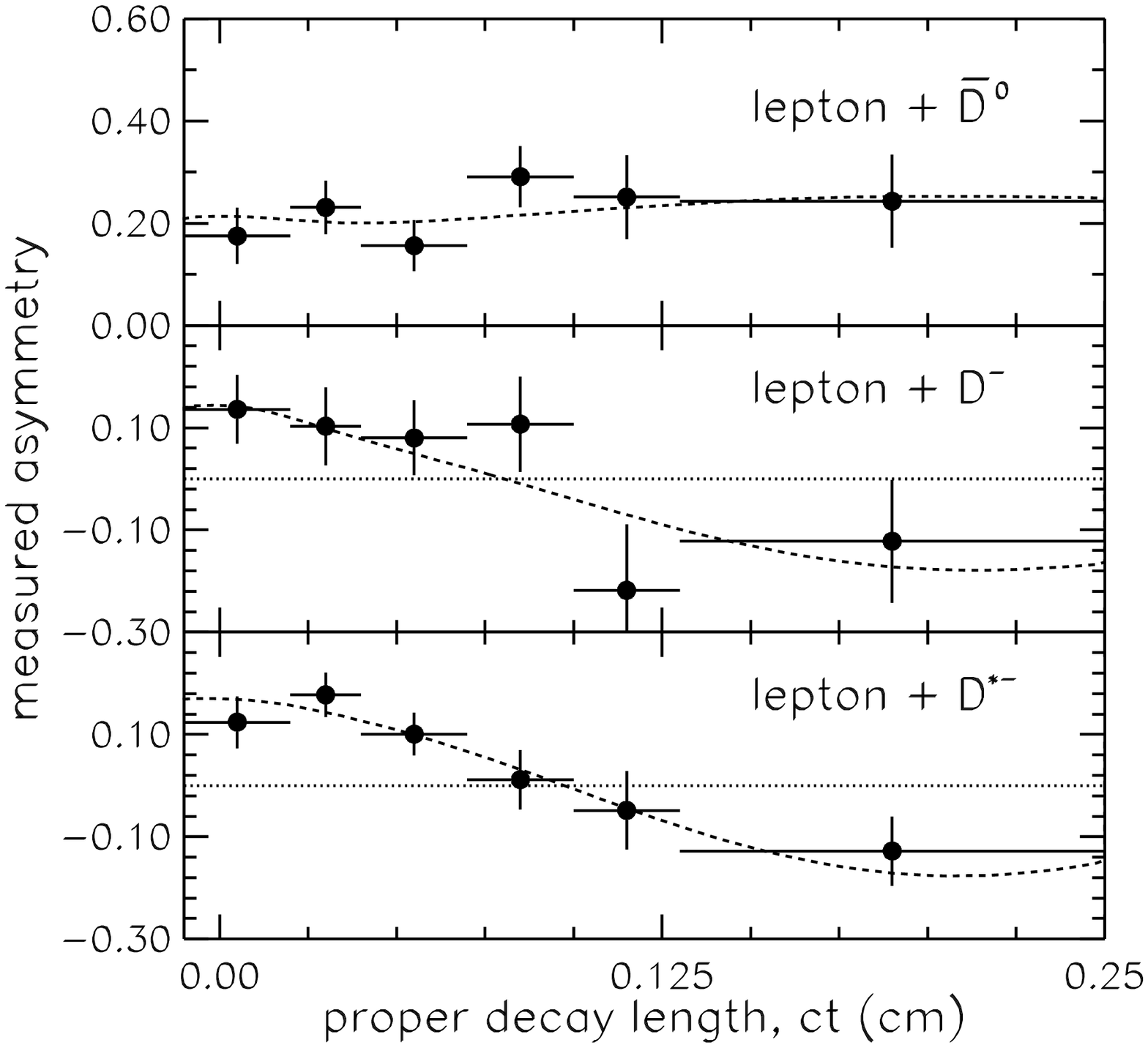}
\epsfysize=2.5in
\epsffile[25 175 510 520]{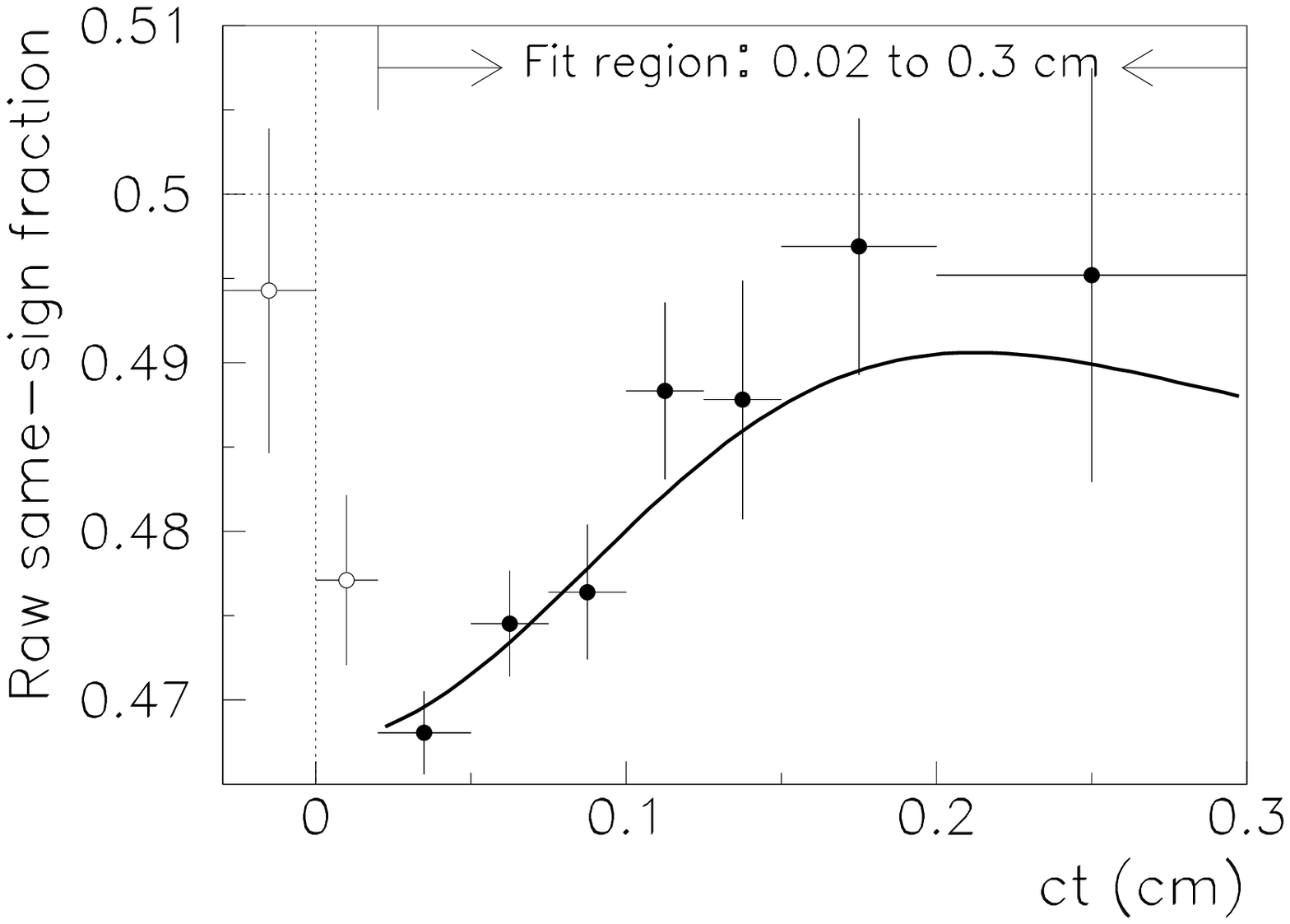}
}
\caption{(a) 
Measured asymmetries as a function of proper decay length using a same side
tag for the
decay signatures: $\bar D^0\ell^+$ (top), $D^- \ell^+$ (middle), and
$D^{*-} \ell^+$ (bottom). 
(b) Fraction of mixed events
as a function of proper decay length using a jet charge and lepton
flavor tag.}
\label{fig:cdfmix}
\end{center}
\end{figure}

$B$~flavor tagging is the crucial element for a $CP$~violation measurement
at the Tevatron. The figure of merit quantifying how well a flavor tagging
algorithm works is the so-called effective tagging efficiency 
$\varepsilon {\cal D}^2$.
Here, $\varepsilon$ is the efficiency for obtaining a particular flavor
tag, and $\cal D$ is the dilution defined by the number of
right tags ($N_R$) and the number of wrong tags ($N_W$):
${\cal D} = (N_R - N_W) / (N_R + N_W)$. CDF determined the tagging power of
various tagging methods with data measuring the time dependence of
$B^0\bar B^0$ flavor oscillations. Such a measurement serves as a
demonstration that a particular flavor tag does work in a hadron collider
environment and determines its $\varepsilon {\cal D}^2$.
Figure~\ref{fig:cdfmix}(a) shows the 
measured mixing asymmetries as a function of proper decay length using a
same side tag~\cite{Abe:1999_1,Abe:1998}. From this measurement, CDF extracts 
$\Delta m_d = (0.471 ^{+0.078}_{-0.068} \pm 0.034)\ \rm ps^{-1}$
and the effective tagging efficiency for the same side tag to be
$\varepsilon {\cal D}^2=(1.8\pm0.4\pm0.3)\%$. As another example, 
Fig.~\ref{fig:cdfmix}(b) shows the fraction of mixed events as a function
of proper decay length using a jet charge and lepton 
flavor tag~\cite{Abe:1999_2}. This measurement yields 
$\Delta m_d = (0.500\pm0.052\pm0.043)\ \rm ps^{-1}$ as well as 
$\varepsilon {\cal D}^2=(0.91\pm0.10\pm0.11)\%$ and
$\varepsilon {\cal D}^2=(0.78\pm0.12\pm0.08)\%$ for a
lepton tag and jet charge tag, respectively.

Applying all three tagging methods to CDF's sample of $J/\psi K^0_S$
events gives the $CP$~asymmetry distribution shown in
Fig.~\ref{fig:cdfsin2b}(b). The data prefer a positive asymmetry, 
resulting in a measurement of 
$\sin 2\beta = 0.79 \pm 0.39 \pm 0.16$. This can be translated into a limit 
on $\sin 2\beta$ being positive ($0<\sin2\beta<1$) at 93\% confidence
level. This is the best direct measurement of $CP$~violation in the
$B$~system to date. With this result, CDF demonstrated that a $CP$~violation
measurement is feasible at the Tevatron in Run\,II. Returning to our initial
question, we think the Tevatron does qualify as a $B$~factory capable of
measuring $CP$~violation.

\section{The Tevatron in Run\,II}

The most important element of the Tevatron upgrade
for Run\,II is the Main Injector. It is a new 150~GeV accelerator, half the
circumference of the Tevatron, which will increase the
antiproton intensity into the Tevatron, providing 20 times higher
luminosities. The Main Injector project was finished in June 1999 with the
first beam circulating at that time.

Run\,II of the Tevatron had originally been defined as 2~fb$^{-1}$ being
delivered in two years to the collider experiments.
Run\,II has recently been extended beyond the initial two
years, to continue until 2006 with no major shutdowns,
maximizing the delivered luminosity to a total of up to 15~fb$^{-1}$.
The current Fermilab schedule fixes the start of Run\,II in March 2001.

\subsection{D\O\ Detector Upgrade}

\begin{figure}[tb]
\begin{center}
\centerline{
\epsfysize=2.6in
\epsffile[20 110 575 500]{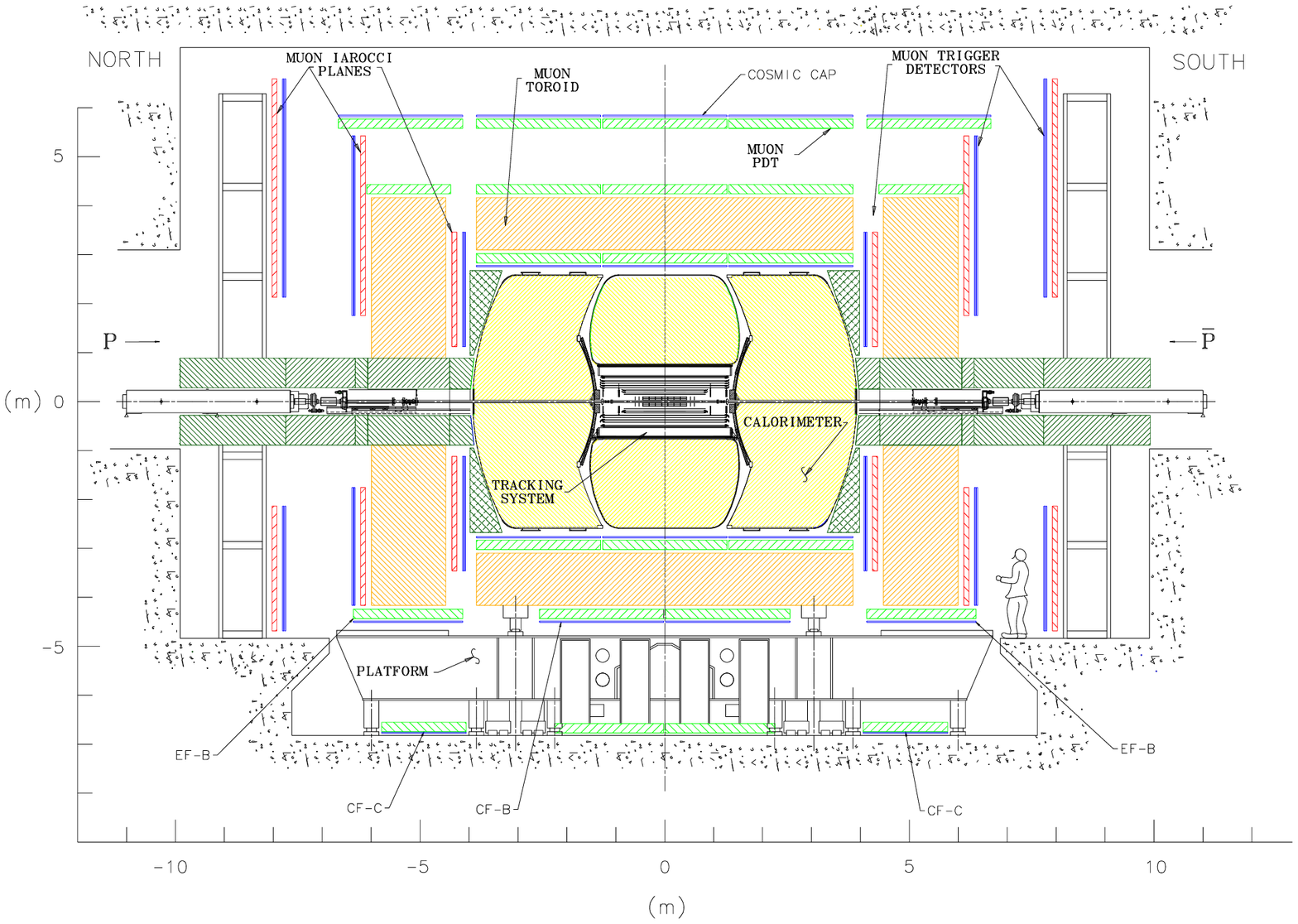}
\epsfysize=2.6in
\epsffile[30 80 570 670]{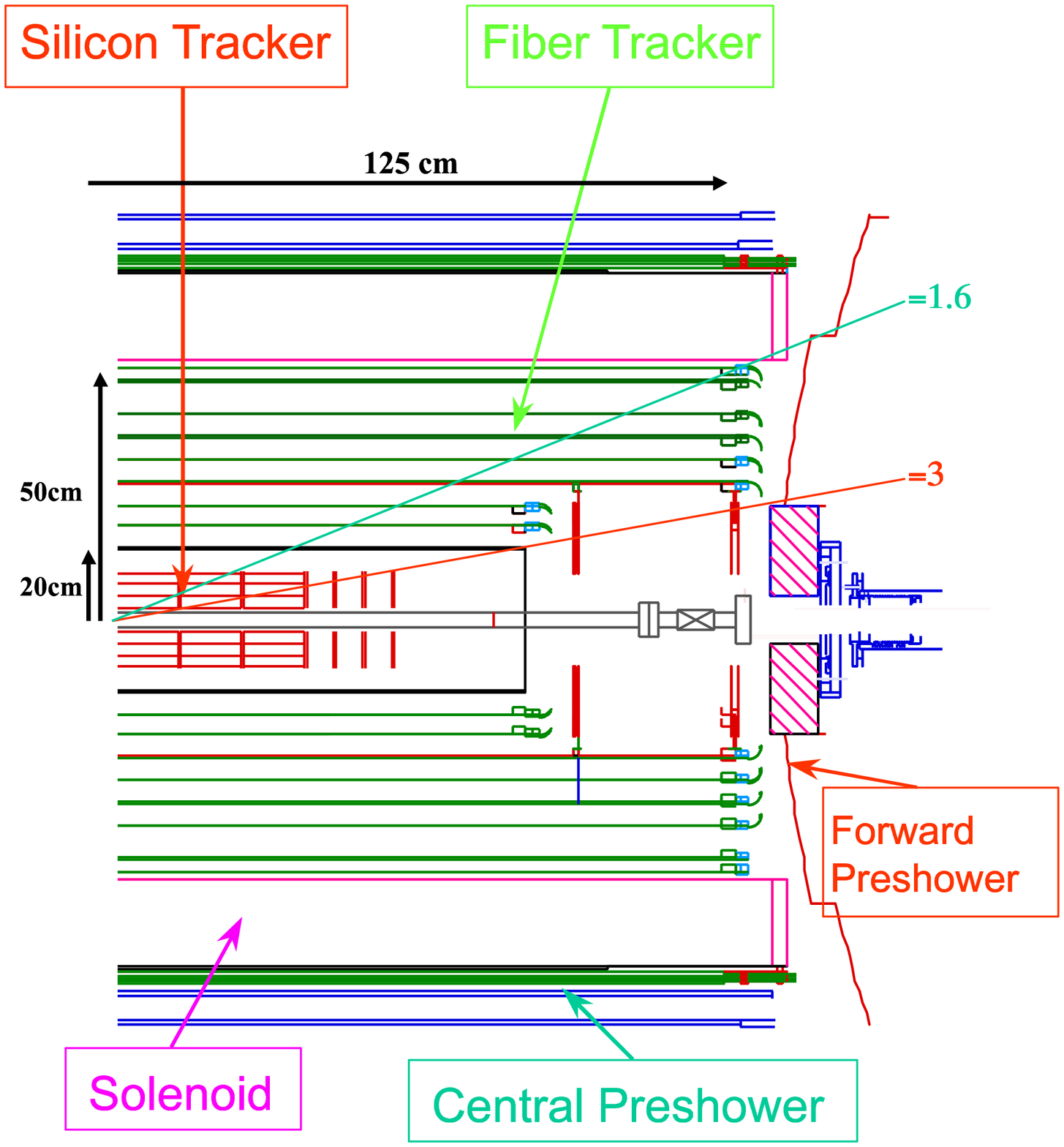}
}
\caption{ 
Left: Cross section of the upgraded D\O\ detector. Right: Longitudinal view
of the D\O\ tracking system.}
\label{fig:d0upgrade}
\end{center}
\end{figure}

The D\O\ detector upgrade is built on previous strengths, combining
excellent calorimetry with good muon coverage and purity. A 
cross section of the upgraded D\O\ detector is shown in
Fig.~\ref{fig:d0upgrade}. 
The most important improvement is a superconducting solenoid ($B=2$~T)
providing significantly improved tracking capabilities (see right hand side
of Fig.~\ref{fig:d0upgrade}). A central fiber
tracker consisting of eight superlayers of scintillating fibers allows a
measurement of the charged particle momentum. Together with the silicon
micro-strip tracker, a momentum resolution of $\sigma(p_T)/p_T = 0.002\,p_T$
will be achieved. The silicon tracker consists of 
six barrels with four layers each ($r\phi$ and $rz$ readout) and
12+4 forward disks reaching out to 1.25\,m in $z$. 
In addition, improvements to the muon system will allow enhanced muon
triggering for $p_T > 1.5$~GeV$/c$ ($|\eta|<2$). Central and forward
preshower detectors will improve electron identification 
and triggering on electrons with $p_T > 1$~GeV$/c$ ($|\eta|<2.5$).
Finally, an impact parameter trigger detecting tracks from displaced
vertices is under development. 

The current D\O\ schedule expects the central and forward preshower
fabrication as well as the central fiber tracker project to be completed by
June 2000. The silicon tracker will be finished by September 2000 and the
full tracking system installed and hooked up a few weeks later. The muon
system will be in place by November 2000 and the calorimeter electronics by
January 2001, allowing D\O\ to be rolled in and ready for beam by February
2001. 

\subsection{CDF Detector Upgrade}

\begin{figure}[tb]
\begin{center}
\centerline{
\epsfxsize=3.3in
\epsffile[1 1 685 465]{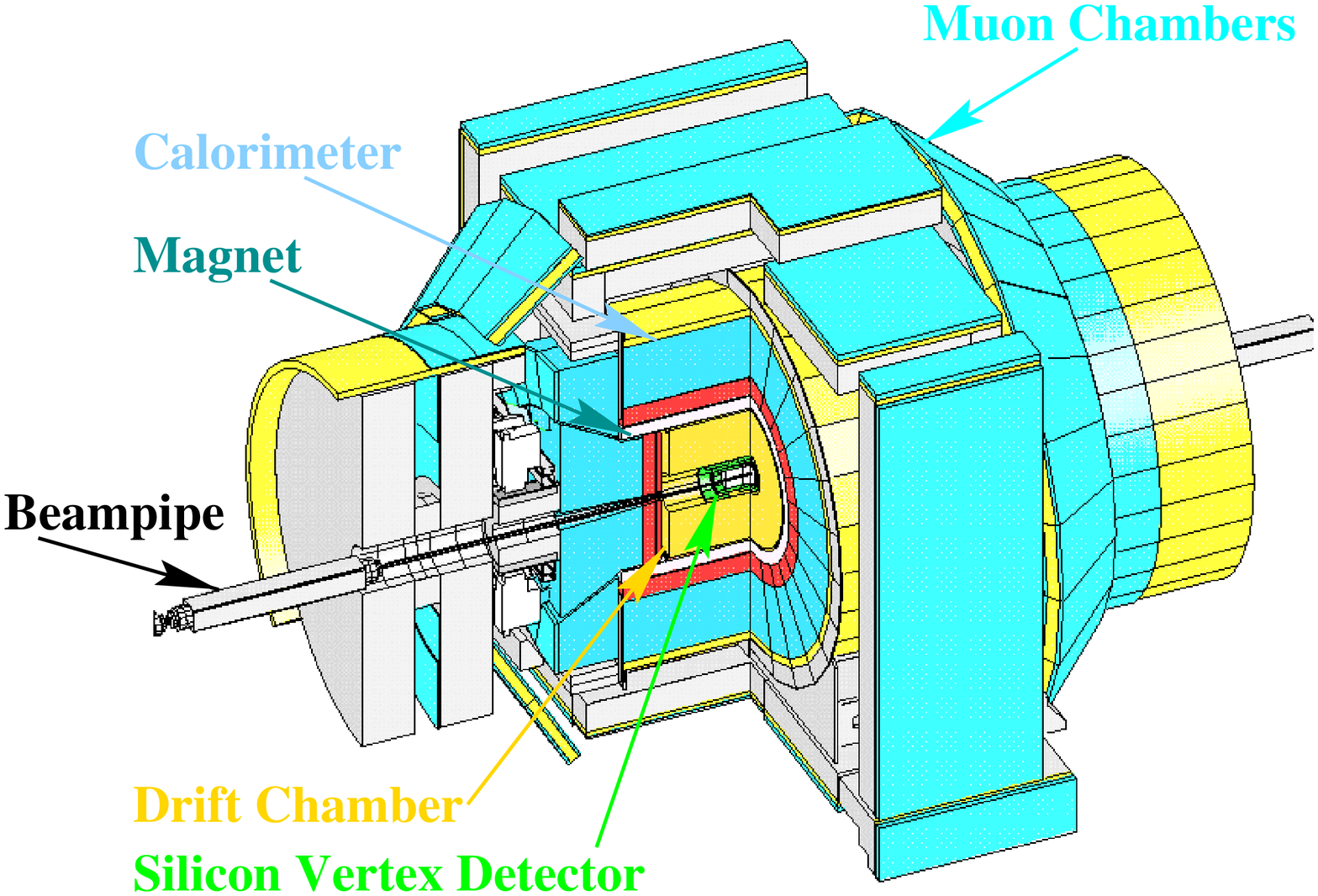}
\epsfxsize=2.7in
\epsffile[40 315 570 730]{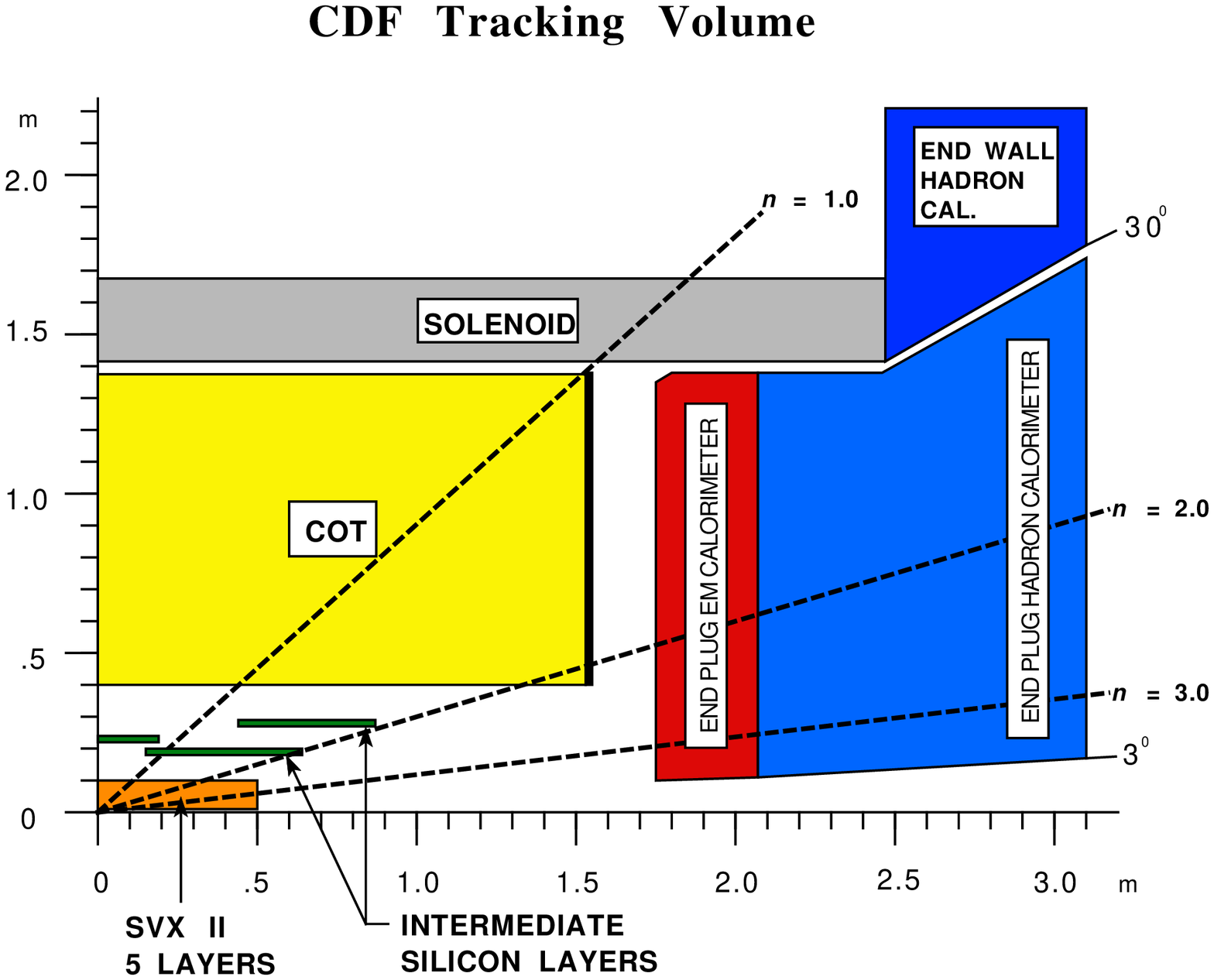}
}
\caption{ 
Left: Schematic cut-away view of the CDF\,II detector. 
Right: Longitudinal view of the upgraded CDF tracking system.}
\label{fig:cdfupgrade}
\end{center}
\end{figure}

The goal for the CDF detector upgrade is to maintain detector
occupancies at Run\,I levels, although many of the detector changes
also provide qualitatively improved detector capabilities.
A schematic view of the CDF\,II detector is shown in
Fig.~\ref{fig:cdfupgrade}. 
One major improvement is to the charged particle tracking system (see
Fig.~\ref{fig:cdfupgrade}),
vital for the $B$~physics program at CDF. 
A new silicon vertex detector will consist of five layers of
double sided silicon from radii of 2.9~cm to 10~cm. The silicon detector
will include three modules covering the entire $p\bar p$~luminous 
region. In addition, an intermediate silicon layer consisting of two
double-sided silicon sensors at larger radii permits stand-alone
silicon tracking out to $|\eta|=2$. 
A new open cell drift chamber (COT) will operate at a beam crossing time
of 132~ns with a maximum drift time of $\sim\!100$~ns.
The COT consists of 96 layers arranged in four axial and four stereo
superlayers. It also provides d$E$/d$x$ information for particle
identification. 

The upgrades to the muon system almost double the central
muon coverage.  A new scintillating tile plug
calorimeter will give good electron identification up to $|\eta|=2$.
New front-end electronics will be installed, and a DAQ upgrade will allow the
operation of a pipelined trigger system.
Finally, two additional upgrade projects 
significantly enhancing the $B$~physics capabilities of the
CDF\,II detector have been approved. These include the installation of a
low-mass radiation 
hard single-sided silicon detector with axial strips at very small radius of
$\sim\!1.6$~cm, as well as the installation of a
time-of-flight system employing 216 three-meter-long
scintillator bars located between the outer radius of the COT
and the superconducting solenoid.

The current CDF schedule foresees cosmic ray running of the detector at the
beginning of 2000 and expects the COT to be installed in April 2000.
A commissioning run will take place from August to
November of 2000. The silicon upgrade will be complete by September 2000
and installed by January 2001. The full CDF\,II detector will be ready for
collisions by March 2001.

\section{Run\,II \boldmath{$B$}~Physics Prospects}

When discussing the  Run\,II $B$~physics prospects in this section, we will
refer to a data sample of 2~fb$^{-1}$ delivered in two years. We
will focus on the prospects for the CDF experiment. D\O\ has similar
expectations. 

\begin{table}[tb]
\begin{center}
\begin{tabular}{l|cc}
\hline
 Flavor tag & $\varepsilon {\cal D}^2$ Run\,I &
 $\varepsilon {\cal D}^2$ Run\,II \\
\hline
\hline
Same side tag & $(1.8\pm0.4\pm0.3)\%$ & 2.0\% \\
Jet charge tag & $(0.78 \pm 0.12 \pm 0.08)\%$ & 3.0\% \\
Lepton tag & $(0.91 \pm 0.10 \pm 0.11)\%$ & 1.7\% \\
Kaon tag & -- & 2.4\% \\
\hline
\end{tabular}
\caption{Summary of effective $B$~flavor tagging efficiencies $\varepsilon
{\cal D}^2$ measured at CDF in Run\,I and corresponding projections for
Run\,II.}
\label{tab:tag}
\end{center}
\end{table}

\begin{figure}[tb]
\begin{center}
\centerline{
\epsfxsize=4.0in
\epsffile[20 55 530 320]{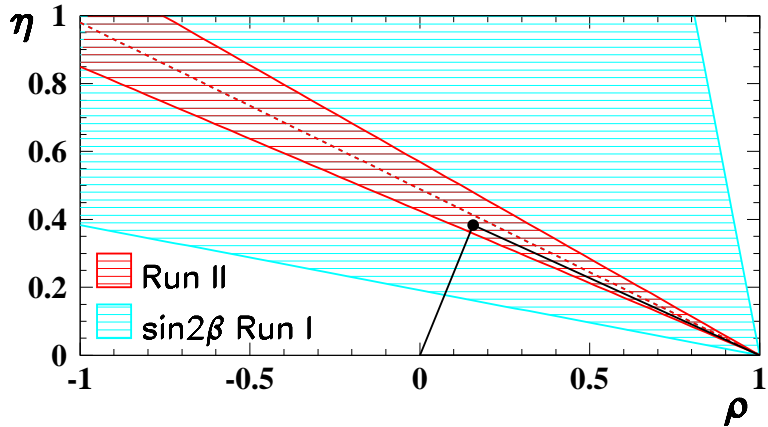}}
\caption{Illustration of the current CDF result
on $\sin 2\beta$ in the $(\rho,\eta)$-plane (light shaded area) and 
expected improvement in Run\,II (dark shaded area).}
\label{fig:sin2bprosp}
\end{center}
\end{figure}

For a measurement of $\sin 2\beta$, CDF expects 10,000 $J/\psi K^0_S$
events in 2~fb$^{-1}$ with the $J/\psi$ decaying to muon pairs and 
$K^0_S \rightarrow \pi^+\pi^-$. With the enhanced tracking and vertexing
capabilities, extended lepton coverage and better particle identification, 
CDF will improve the effective
flavor tagging efficiencies for the different taggers as detailed in
Table~\ref{tab:tag} to a total $\varepsilon {\cal D}^2$ of approximately
9.1\%. With this, CDF expects to measure $\sin 2\beta$ with an uncertainty
of $\sim\!0.07$. Figure~\ref{fig:sin2bprosp} shows the current CDF result
on $\sin 2\beta$ in the $(\rho,\eta)$-plane where the light shaded area
indicates the present $1\sigma$ uncertainty. To illustrate the improvements in
Run\,II, the dark shaded area displays the expected error on $\sin2\beta$
in Run\,II with 2~fb$^{-1}$.  

With respect to other $CP$~modes, CDF plans to measure the time dependence
of the $CP$~asymmetry in $B^0 \rightarrow \pi^+\pi^-$ decays, determining
$\sin 2\alpha$. CDF will use a displaced track trigger which will trigger
on hadronic tracks from long-lived particles such as $B$~hadrons. With a
fast track trigger at Level~1, CDF finds track pairs in the COT
with $p_T$ greater than 1.5~GeV$/c$. At Level~2, these tracks are
linked into the silicon vertex detector, and cuts on the track impact
parameter $d > 100$ $\mu$m ($\sigma(d) \sim 25\ \mu$m) are applied. 
CDF expects to collect 4000-7000
$B\rightarrow \pi\pi$ events, assuming ${\cal B}(B^0 \rightarrow \pi^+\pi^-) = 
(4.7^{+1.8}_{-1.5}\pm0.6)\cdot 10^{-6}$ as measured by CLEO~\cite{Poling}. In
addition, there will be about four times more $B \rightarrow K \pi$
decays. The implications of this background for an extraction of $\sin
2\alpha$ at CDF are still under study. 

For a measurement of $\sin\gamma$ in Run\,II, CDF can use 
$B_{\mbox{\scriptsize S}}^0$~mesons. 
A signal of about 700 
$B_{\mbox{\scriptsize S}}^0 / \bar B_{\mbox{\scriptsize S}}^0 
\rightarrow D_{\mbox{\scriptsize S}}^{\pm}K^{\mp}$ events
is expected. This might allow for an initial measurement of $\sin\gamma$
with 2~fb$^{-1}$ in Run\,II.

$B^0 \bar B^0$ and
$B_{\mbox{\scriptsize S}}^0 \bar B_{\mbox{\scriptsize S}}^0$ flavor
oscillations measure the Cabibbo-Kobayashi-Maskawa matrix elements
$|V_{td}|/|V_{ts}|$. 
The recently approved detector
upgrades play an important role in CDF's prospects for measuring
$B_{\mbox{\scriptsize S}}^0$ mixing.  
The additional inner layer of silicon
improves the time resolution from 
$\sigma_t = 0.060$~ps to 0.045~ps. This will be important if 
$\Delta m_{\mbox{\scriptsize S}}$ is unexpectedly large. The time-of-flight
system will enhance the effectiveness of $B$~flavor tagging, especially
through same side tagging with kaons and opposite side kaon tagging, to a
total $\varepsilon {\cal D}^2 \sim 11.3\%$. CDF expects a signal of
15,000--23,000 fully reconstructed 
$B_{\mbox{\scriptsize S}}^0 \rightarrow 
D_{\mbox{\scriptsize S}}^- \pi^+,\ 
D_{\mbox{\scriptsize S}}^- \pi^+\pi^-\pi^+$
events from the two-track hadronic trigger in 2~fb$^{-1}$. For 20,000 
$B_{\mbox{\scriptsize S}}^0$ events, a $5\sigma$ measurement of 
$\Delta m_{\mbox{\scriptsize S}}$ will be possible at CDF for 
$\Delta m_{\mbox{\scriptsize S}}$ values up to 40~ps$^{-1}$. The
current limit on $\Delta m_{\mbox{\scriptsize S}}$ is
14.3~ps$^{-1}$~\cite{Blaylock} at 95\% C.L.  It is noteworthy that physics with 
$B_{\mbox{\scriptsize S}}^0$ mesons will be unique to the Tevatron until
the turn-on of the LHC in 2006. 

\section{Conclusions}

The CDF and D\O\ detector upgrades are well under way with data taking 
starting in March 2001. There are excellent prospects for $B$~physics in
Run\,II, allowing a measurement of $\sin 2\beta$ with an uncertainty of 0.07.
A discovery of $B_{\mbox{\scriptsize S}}^0$ mixing is possible for 
$\Delta m_{\mbox{\scriptsize S}}$ values up to 40~ps$^{-1}$. The extension
of Run\,II until 2006 will further increase the sensitivity and the
$B$~physics potential at the Tevatron. CDF and D\O\ are looking forward to
joining the party with the $B$~factories.

\def\Discussion{
\setlength{\parskip}{0.3cm}\setlength{\parindent}{0.0cm}
     \bigskip\bigskip      {\Large {\bf Discussion}} \bigskip}
\def\speaker#1{{\bf #1:}\ }

\Discussion

\speaker{Michail Danilov (ITEP, Moscow)}
What is the sensitivity of the D\O\ experiment for $B$~physics studies in
Run\,II? 

\speaker{Paulini} 
As mentioned in my presentation, the prospects for D\O\ are similar to the
ones at CDF. D\O\ expects about 8500 $J/\psi K^0_S$ events with $J/\psi
\rightarrow \mu^+\mu^-$ but they plan to also trigger on $e^+e^-$ pairs
resulting in additional 6500 $J/\psi K^0_S$ events. From a time-dependent
analysis, D\O\ expects to measure $\sin2\beta$ with an uncertainty of
0.07. In addition, an impact parameter trigger project has recently been
approved allowing D\O\ to detect $B\rightarrow \pi\pi$ events and also to
explore $B_{\mbox{\scriptsize S}}^0 \bar B_{\mbox{\scriptsize S}}^0$ mixing.

\end{document}